\documentclass[sigconf]{acmart}

\usepackage{amsmath,amsfonts}
\usepackage{algorithmicx}
\usepackage{graphicx}
\usepackage{textcomp}
\usepackage{xcolor}

\def\BibTeX{{\rm B\kern-.05em{\sc i\kern-.025em b}\kern-.08em
    T\kern-.1667em\lower.7ex\hbox{E}\kern-.125emX}}
\usepackage{booktabs} 
\usepackage{tabularx}
\usepackage{array}
\usepackage{subcaption}
\usepackage{multicol}
\newcolumntype{$}{>{\global\let\currentrowstyle\relax}}
\newcolumntype{^}{>{\currentrowstyle}}

\setcopyright{none} 

\begin{document}

\title{\huge{A Unified Model for Gate Level Propagation Analysis}}

\settopmatter{printacmref=false} 
\renewcommand\footnotetextcopyrightpermission[1]{} 
\pagestyle{plain}

\author{Jeremy Blackstone}
\affiliation{University of California, San Diego, USA}
\email{jblackst@ucsd.edu}

\author{Wei Hu}
\affiliation{Northwestern Polytechnical University, China}
\email{weihu@nwpu.edu.cn }

\author{Armaiti Ardeshiricham}
\affiliation{University of California, San Diego, USA}
\email{aardeshi@eng.ucsd.edu}

\author{Lu Zhang}
\affiliation{Northwestern Polytechnical University, China}
\email{willvsnick@nwpu.edu.cn }

\author{Alric Althoff}
\affiliation{University of California, San Diego, USA}
\email{aalthoff@eng.ucsd.edu}

\author{Ryan Kastner}
\affiliation{University of California, San Diego, USA}
\email{kastner@ucsd.edu}

 \begin{abstract}
Classic hardware verification techniques (e.g., X-propagation and fault-propagation) and more recent hardware security verification techniques based on information flow tracking (IFT) aim to understand how information passes, affects, and otherwise modifies a circuit. These techniques all have separate usage scenarios, but when dissected into their core functionality, they relate in a fundamental manner. In this paper, we develop a common framework for gate level propagation analysis. We use our model to generate synthesizable propagation logic to use in standard EDA tools. To justify our model, we prove that Precise Hardware IFT is equivalent to gate level X-propagation and imprecise fault propagation. We also show that the difference between Precise Hardware IFT and fault propagation is not significant for 74X-series and '85 ISCAS  benchmarks with more than 313 gates and the difference between imprecise hardware IFT and Precise Hardware IFT is almost always significant regardless of size.
\end{abstract}
 \maketitle




\section{Introduction}
A variety of disciplines track the movement of data through hardware designs in order to make important design decisions. For example, information flow tracking (IFT) monitors the propagation of security critical information to determine compliance to security properties~\cite{Imprecise_Sec, Specific_GLIFT, GLIFT_Complexity, GLIFT, First_GLIFT}; X-propagation aims to understand how non-deterministic values originating from uninitialized registers influence reliability and where they could possibly flow to~\cite{Xprop, LoveX, XPA_Sim, Accurate_XPA, Trojan_XPA}; and fault propagation models the effect of faulty values on system reliability~\cite{TSV, TMR1, TMR2} and susceptibility to implementation attacks~\cite{TVVF, SSE}.

Although these techniques seek to provide designers with very different types of insight about a design, the steps taken to discover such information and their uninterpreted results are very similar and in some cases identical. By taking advantage of these similarities, it is possible to cast these problems into a unified formal model for gate level propagation analysis. In this work, we develop this unified propagation model, describe how techniques from fault propagation, IFT, and X-propagation fit into this model, and analytically and quantitatively describe the differences between these techniques. 
\begin{figure}[!t]
\centering
\includegraphics[width=.6\columnwidth]{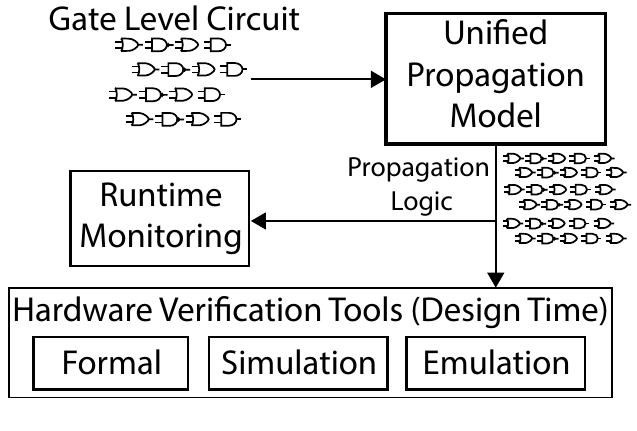}%
\caption{\textbf{Our unified propagation model takes a gate level circuit as input and generates a separate synthesizeable propagation logic. This propagation logic can be analyzed during design time using existing hardware verification tools to perform formal, simulation, and emulation. If it passes verification, the propagation logic is no longer needed; it is discarded and the original circuit goes through the traditional synthesis process. Additionally, because the propagation logic is itself synthesizeable, it can be fabricated as a runtime monitor checking for propagation violations in the original circuit. }}
\label{fig:framework}
\vspace{-4mm}
\end{figure}

Our model takes the original circuit as input and formalizes separate analysis logic called propagation logic that tracks the propagation of values in hardware designs as shown in Figure \ref{fig:framework}. This logic processes labels alongside the original circuit inputs to mark some data that we wish to track. The main benefit of our propagation logic is that it is synthesizable. This allows us to analyze the logic using existing hardware design tools for verification of different propagation properties such as security violations, initialization errors, and faulty data. For example, we could use formal verification tools to prove that two modules are isolated from each other. Using existing hardware simulation tools we could run testbenches to see if a particular input affects a particular output. Furthermore, we could use existing hardware emulation tools to perform verification for larger hardware designs. All three of these techniques can be done at design time so the propagation logic does not need to be implemented providing a zero overhead verification technique for propagation analysis of hardware designs. Additionally, the propagation logic can be realized into its own hardware circuit performing analysis alongside the original circuit for which information is being tracked. 

The major contributions of this paper are:
\begin{itemize}
\item{Developing a general framework for performing symbolic execution that provides synthesizable analysis logic for multiple data propagation-related problems;}
\item{Developing propagation logic to track transient faults;}
\item{Quantifying the precision differences between imprecise hardware IFT, Precise Hardware IFT, and fault propagation.}
\end{itemize}

The remainder of this paper is organized as follows. In Section \ref{sec:preliminary} we show how using our unified propagation model allows designers to encode different types of propagation analysis techniques using propagation logic and labels.  Section \ref{sec:comparison} provides an analytical comparison of the techniques to show similarities and differences and Section \ref{sec:results} makes a quantitative comparison of the techniques to highlight how much the techniques differ and how different circuits affect the amount of data propagation. Section \ref{sec:related} discusses related work and Section \ref{sec:conclusion} concludes the paper. 

\section{Unified Propagation Model}

This section formalizes our unified propagation model. We describe how to use attributes to build different types propagation logic, e.g. related to information flow tracking, faults, and x-propagation. And we describe the problem of precision in the propagation models including the benefits of safe but imprecise propagation logic.

\label{sec:preliminary}
\subsection{Circuit Model}
In digital hardware design, the most frequently used circuit models are Boolean functions and Boolean gates. Equation (\ref{eqn:circuit}) is a general representation of a Boolean circuit, where $I_1, I_2, \dots, I_n$ are the inputs while $O$ is the output.
\begin{equation}
\label{eqn:circuit}
    O = f(I_1, I_2, \cdots, I_n)
\end{equation}

 In this work, we focus our analysis on gate level netlists and thus employ Boolean gates as our circuit model. The Boolean circuits are composed of Boolean operators or primitive gates, e.g., AND, OR, XOR and Invertor. 

\subsection{Attribute Label}
The key idea for our propagation model is the addition of attributes or labels to the Boolean values that indicate something related to the propagation, e.g., taint, X-state, and fault. By observing the attribute labels, we can reason about important design properties related to security, resilience, and fault tolerance. Table \ref{tab:attribute} shows how attributes are encoded under different application scenarios.
\begin{table}[!ht]
  \caption{Attribute encoding for different applications.}
  \centering
  \label{tab:attribute}
  \begin{tabular}{ccc}
    \toprule
    Application & 1-Label & 0-Label \\
    \midrule
    IFT & Tainted & Untainted  \\ 
    XPA & Non-deterministic  & Deterministic  \\ 
    FPA & Faulty  & Correct  \\ 
  \bottomrule
\end{tabular}
\vspace{-3mm}
\end{table}

\textbf{Definition 1 - Taint:} In hardware IFT, we associate data with a security label called \emph{taint}. We define a data object to be tainted if the data object contains secret information in confidentiality analysis or if it carries untrusted information in integrity analysis~\cite{GLIFT}. Taint propagation decides whether or not it is possible for tainted inputs to affect an output. If there is any scenario where a tainted value can cause a change in the output, then the output is marked as tainted. Otherwise, the output is untainted and there is no tainted information flow.

\textbf{Definition 2 - Non-deterministic:} If there is any scenario where the X-state can cause a change in the output, then the output is \emph{non-deterministic}. If it is not possible for the X-state to cause a change in the output, then the output is  \emph{deterministic}. We define an output data bit to be  \emph{non-deterministic} if toggling the logical value of a particular unknown input changes the output~\cite{GLIFT}. 

\textbf{Definition 3 - Faulty:} If there is any scenario where a misinterpreted logic value can cause a change in the output, then the output is faulty. If it is not possible for a misinterpreted logic value to cause a change in the output, then the output is correct. We define an output data bit to be faulty if simultaneously flipping the logical value of all faulty inputs changes the output \cite{GLIFT}.

\subsection{Precision}
Depending on the attribute label, it is possible to have different rules for deriving the propagation logic~\cite{Imprecise_Sec}. This creates different propagation logic with potentially varying precision and complexity. The precision of propagation logic is defined as the degree to which propagation logic models real circuit behavior. A precise propagation logic updates the attribute label of the output \emph{if and only if} at least one input with the same attribute label has an effect on the output. By comparison, an imprecise propagation logic may contain false positives or false negatives. A false positive happens when the precise propagation logic sets the output attribute label to logic `0' while the imprecise propagation logic sets the label to logic `1'. A false negative is a case when the precise propagation logic sets the output attribute label to logic `1' while the imprecise propagation logic sets the label to logic `0'. A false positive indicates non-existent circuit behavior, which makes the propagation logic conservative when modeling circuit behavior. A false negative fails to capture a circuit behavior that does exist, which makes the propagation logic erroneous. It is totally safe (although not ideal) for propagation logic to have false positives while false negatives should be never allowed.

A safe, imprecise propagation logic can ignore the effect that certain inputs have on the output when calculating the attribute label for the output to reduce the computational cost. However, this process introduces false positives. More formally, an imprecise propagation logic is derived from Equation (\ref{eqn:shadow}) by reducing some inputs while still keeping their security labels. A frequently used version of imprecise propagation logic can be described as Equation (\ref{eqn:imprecise}), which completely ignores the effect of inputs on attribute label propagation.
\begin{equation}
\label{eqn:imprecise}
    L(O) = f_{Label}(L(I_1),\ L(I_2), \cdots,\ L(I_n))
\end{equation}

One way to define precise propagation logic is enumerating propagation logic truth tables. These truth tables are capable of analyzing all possible combinations of input values and attribute labels to determine the output label for all scenarios. While this is feasible for small primitive gates, such adjective enumeration is usually impractical for larger logic functions. For large circuits, it is far more efficient to constructively generate the propagation logic for the circuit from smaller propagation logic functions for primitive gates in a manner similar to technology mapping \cite{GLIFT_Complexity}.

\subsection{Propagation Logic}
Propagation logic determines how labeled inputs propagate to the output labels. That is, it creates logical rules for the output attribute labels based on the input values, their attribute labels, and the method of propagation. The function is defined as a map from the circuit inputs and their attribute labels to the attribute label of the output as shown in Equation (\ref{eqn:shadow}). $L(I_1), L(I_2), \cdots, L(I_n)$ and $L(O)$ are the attribute labels of inputs $I_1, I_2, \cdots, I_n$ and $O$ respectively.
\begin{equation}
\label{eqn:shadow}
    L(O) = f_{Label}(I_1,\ I_2, \cdots,\ I_n,\ L(I_1),\  L(I_2), \cdots,\ L(I_n))
\vspace{-2mm}
\end{equation}

Table \ref{tab:tt_AND} shows a partial truth table for label propagation of the two-input AND, OR and XOR gates, where the $a$ and $b$ columns represent input values for $a$ and $b$; the $a_t$ and $b_t$ columns are their label values and the remaining columns are the output taint values.
\begin{table}[!htp]
  \caption{Partial Truth Table for label Propagation of two-input AND, OR and XOR gates.}
  \centering
  \label{tab:tt_AND}
  \begin{tabular}{ccccccccccl}
    \toprule
    Line&a&b&$a_l$&$b_l$&$AND\_O_l$&$OR\_O_l$&$XOR\_O_l$\\
    \midrule
    1 &	-& -& 1 & 1 & 1 & 1 & 1\\
    2 &	-& -& 0 & 0 & 0 & 0 & 0\\
    3 & \textbf{0} &0 &\textbf{0}&\textbf{1}&\textbf{0}&1&1\\
    4 & 0 &\textbf{1}&\textbf{1}&\textbf{0}&1&\textbf{0}&1\\
  \bottomrule
\end{tabular}
\\[10pt]
\vspace{-4 mm}
\end{table}

Take the two-input AND gate as an example. \emph{Line 1} of Table \ref{tab:tt_AND} shows that when both inputs have a label the output will be always have a label while \emph{line 2} indicates that when both inputs are do not have a label, the output will not have a label.

When input $a$ is `0' without a label, the output will not have a label regardless of the value or label of $b$. This is because the output will be dominated by $a$ to be `0'; the value of $b$ does not have an effect on the output. \emph{Line 3} shows such a case. \emph{Line 4} shows that when one input has a label, the output will be labeled if the other input is logic `1'.
 
 When considering the full truth table, we can derive the propagation logic for all of the gates as shown below.

Propagation logic for a two-input AND gate:

\begin{equation}
\label{eq:GLIFT_AND}
O_l = a \cdot b_l + b \cdot a_l + a_l \cdot b_l \\
\end{equation}

Propagation logic for a two-input OR gate:

\begin{equation}
\label{eq:GLIFT_OR}
O_l = \bar{a} \cdot b_l + \bar{b} \cdot b_l + a_l \cdot b_l \\
\end{equation}

Propagation logic for a two-input XOR gate:

\begin{equation}
\label{eq:GLIFT_XOR}
    O_l = a_l + b_l \\
\end{equation}

Propagation logic for a NOT gate:

\begin{equation}
\label{eq:GLIFT_NOT}
    O_l = i_l \\
\end{equation}

The propagation logic for NAND, NOR and XNOR can be derived from a combination of the propagation logic for NOT followed by that for AND, OR and XOR respectively. According to Equations (\ref{eq:GLIFT_AND}) to (\ref{eq:GLIFT_NOT}), the propagation logic for NAND, NOR, XNOR are identical to those for AND, OR and XOR respectively.

\begin{figure}[!t]
\centering
\includegraphics[width=2 in]{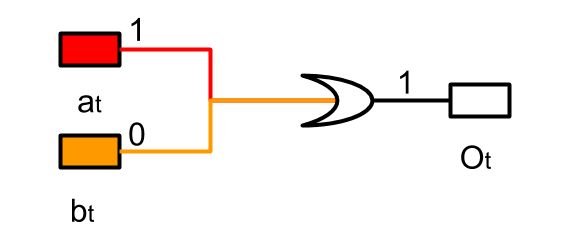}%
\caption{\textbf{Imprecise hardware IFT propagation logic for AND gate}}
\label{fig:CA}
\vspace{-4mm}
\end{figure}

\begin{figure}[!t]
\centering
\includegraphics[width=1.0\columnwidth]{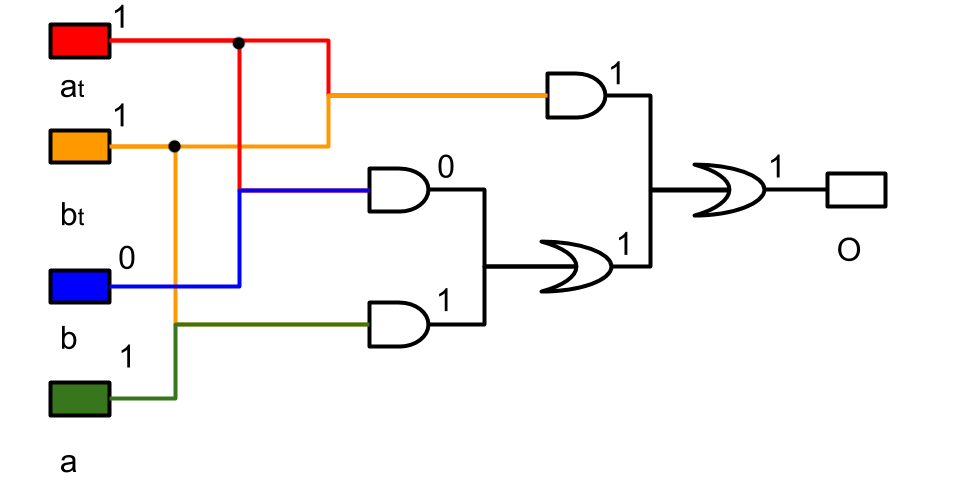}%
\caption{\textbf{Precise hardware IFT propagation logic for AND gate}}
\label{fig:PA}
\vspace{-4mm}
\end{figure}

Figure \ref{fig:CA} shows the propagation logic using imprecise hardware IFT and Figure \ref{fig:PA} is constructed from the propagation logic in Equation (\ref{eq:GLIFT_AND}). While the logic in Figure \ref{fig:CA} propagates the label regardless of the inputs, the logic in Figure \ref{fig:PA} does more precise analysis to validate that it is not possible for input $a$ to have any effect on the output under the given input vector.  We take into account the effect of logic values in label propagation. As a result, intricacies of the logic gates actually prevent some tainted information flows.  

Further, our propagation logic may still contain a certain amount of false positives if the propagation logic is created constructively from those smaller primitive gates. This is due to variable correlation resulting from reconvergent fanouts. However, generating totally precise propagation logic has been proven to be an NP-complete problem \cite{GLIFT_Complexity}. Furthermore, the false positives are not nearly as significant as the taint explosion from imprecise Hardware IFT as we demonstrate in section \ref{sec:diff}. The propagation logic derived in this paper provides a good balance between precision and analysis complexity.

\begin{figure}[!t]
\centering
\includegraphics[width= 3 in]{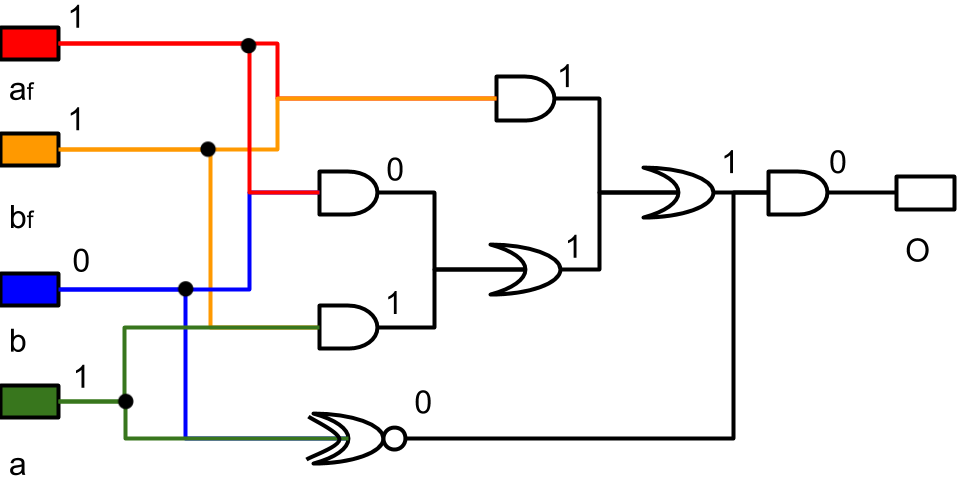}%
\caption{\textbf{Precise FPA propagation logic for AND gate.}}
\label{fig:fPA}
\vspace{-4mm}
\end{figure}

\subsection{Propagation Logic for Fault Propagation}
In a general sense, a fault is any disruption in the normal execution of a device that leads to unexpected results.  However, we focus on tracking transient faults due to their similarities to the taint and non-deterministic labels. A transient fault occurs during execution when the silicon becomes ionized and creates a current that causes the signal value to be incorrectly interpreted by the circuit. This results in a logic value of `0' being interpreted as a `1' or logic value of `1' being interpreted as a `0'. Fault propagation seeks to understand the impact that erroneous behavior will have on a design. It can be used to determine how vulnerable hardware designs are to fault attacks \cite{TVVF} and determine where to apply redundancy to correct the faults and increase the reliability of the device \cite{TMR1}.

While some transient faults may affect the output causing the system to malfunction, others do not affect output and allow the system to function normally. In order to determine which situations produce correct results and which produce faulty results, we construct truth tables and utilize them to derive shadow logic for primitive gates. We refer to this technique as Fault Propagation Analysis (FPA).

\begin{table}
  \caption{Partial Truth Table for Fault Propagation of 2 input AND gate, OR gate and XOR gate.}
  \centering
  \label{tab:f_tt_AND}
  \begin{tabular}{ccccccccccl}
    \toprule
    Line&a&b&$a_ f$&$b_f$&$AND\_O$&$OR\_O$&$XOR\_O$\\
    \midrule
     \textbf{1}&\textbf{0}&\textbf{0}&\textbf{0}&\textbf{0}&\textbf{0}&\textbf{0}&\textbf{0}\\ 
     2&		0&	0&	0&		1&			0	&	1&1\\ 
     \textbf{3}&\textbf{0}&\textbf{0}&\textbf{1}&\textbf{1}&1&1&\textbf{0}\\
     \midrule
     \textbf{4}&\textbf{0}&\textbf{1}&\textbf{0}&\textbf{0}&\textbf{0}&\textbf{1}&\textbf{1}\\ 
     5&		0&	1&	1&		0&			1	&	1&0\\
     \textbf{6}&\textbf{0}&\textbf{1}&\textbf{1}&\textbf{1}&\textbf{0}&\textbf{1}&\textbf{1}\\
     \midrule
     7&		\textbf{1}&	\textbf{1}&\textbf{0}&\textbf{0}&\textbf{1}&\textbf{1}&\textbf{0}\\ 
     8&		1&	1&	0&		1&			0	&	1&1\\
     \textbf{9}&	\textbf{1}&\textbf{1}&\textbf{1}&\textbf{1}&			0	&	0&\textbf{0}\\
  \bottomrule
\end{tabular}
\\[10pt]
\vspace{-8mm}
\end{table}

Table \ref{tab:f_tt_AND}  summarizes when outputs will or will not be faulty for the primitive gates by comparing gates with correct inputs to gates with one of more faulty inputs. The $AND\_O$, $OR\_O$ and $XOR\_O$ columns indicate the outputs of AND, OR and XOR gates respectively. 

The following propagation logic can be used to express the information in the AND gate column of Table \ref{tab:f_tt_AND}
\begin{equation}
\label{eq:FPA_AND}
O_f = (a \cdot b_f + b \cdot a_ f + a_ f \cdot b_f)  \cdot \bar{(a \oplus b)} \\
\end{equation}

The following propagation logic can be used to express the information in the OR gate column of Table \ref{tab:f_tt_AND}
\begin{equation}
\label{eq:FPA_OR}
O_f = (\bar{a} \cdot b_f + \bar{b} \cdot a_f + a_f \cdot b_f) \cdot \bar{(a \oplus b)} \\
\end{equation}

The following propagation logic can be used to express the information in the XOR gate column of Table \ref{tab:f_tt_AND}
\begin{equation}
\label{eq:FPA_XOR}\\
O_f = a_f \oplus b_f
\end{equation}

The following propagation logic can be used to express a NOT gate
\begin{equation}
\label{eq:FPA_NOT}
    O_f = i_f \\
\end{equation}

The propagation logic for NAND, NOR and XNOR can be derived from a combination of the propagation logic for NOT followed by that for AND, OR and XOR respectively. According to Equations (\ref{eq:FPA_AND}) to (\ref{eq:FPA_NOT}), the propagation logic for NAND, NOR, XNOR are identical to those for AND, OR and XOR respectively.

Although most cases are identical to Precise Hardware IFT and GLX, there are two slight differences. If AND gates and OR gates have inputs with complementary logic values that are both faulty, they produce the correct output (Line 6). Additionally, XOR gates will produce the correct output regardless of the input logic values if both inputs are faulty(Lines 3, 6, and 9). We refer to these special cases as fault masking. Furthermore, we refer to FPA that includes fault masking as precise FPA and FPA that does not include fault masking as imprecise FPA.

\begin{figure*}[!t]
\centering
\includegraphics[width= 6.5 in]{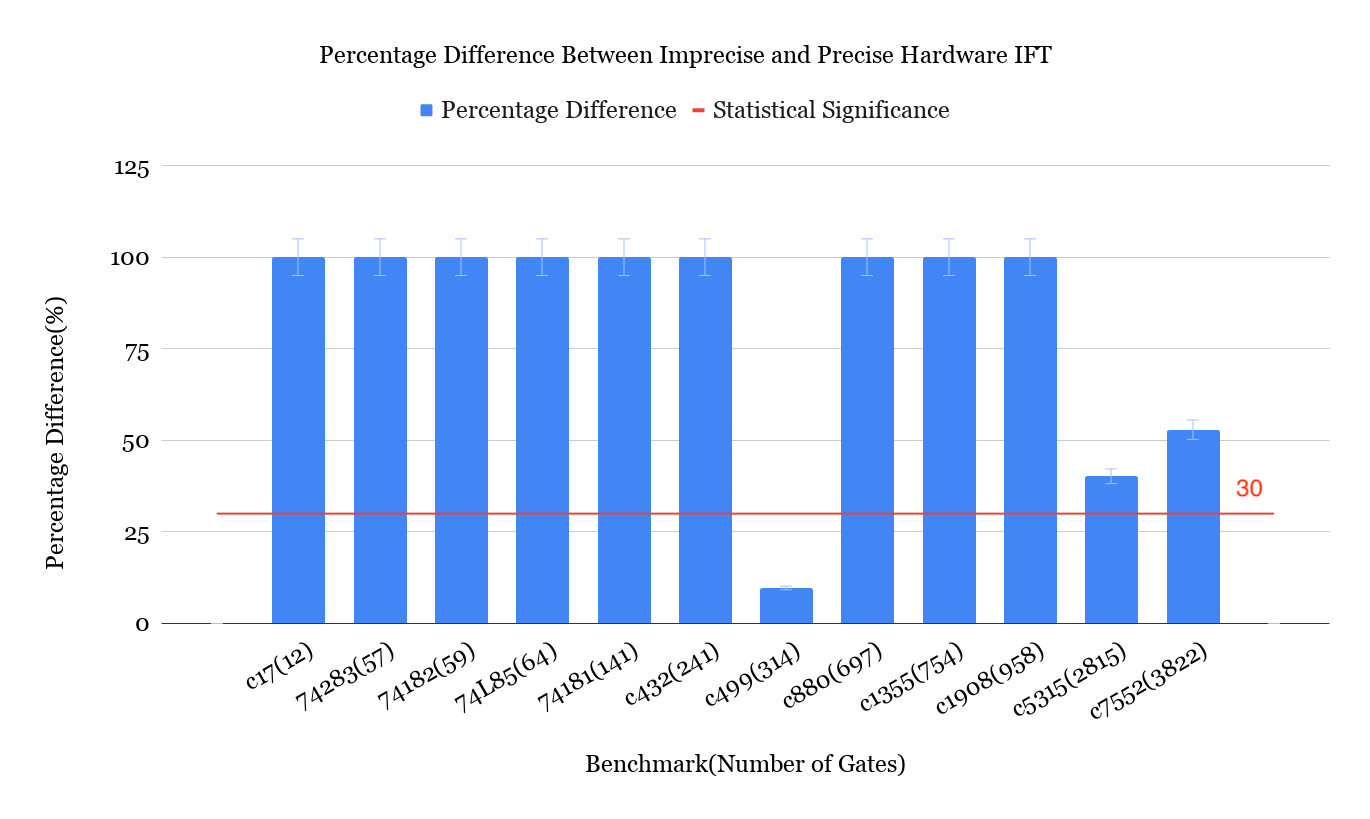}%
\caption{\textbf{The difference between imprecise hardware IFT and Precise Hardware IFT is almost always significant regardless of size.}}
\label{fig:Sig_graph_iGLIFT}
\end{figure*}
\begin{figure*}[!t]
\centering
\includegraphics[width= 6.5 in]{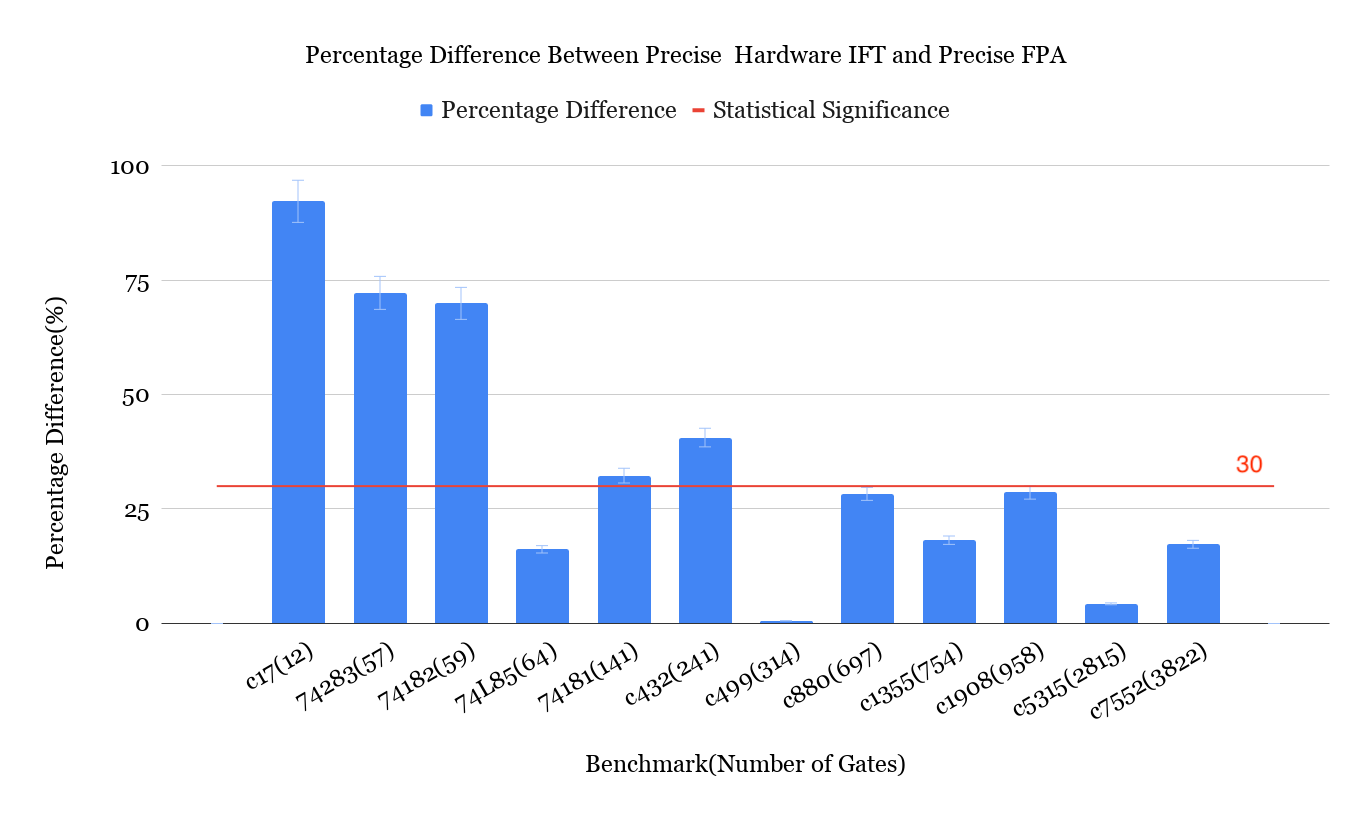}%
\caption{\textbf{The difference between Precise Hardware IFT and FPA is not significant for benchmarks with more than 300 gates.}}
\label{fig:Sig_graph_FPA}
\vspace{-4mm}
\end{figure*}

\begin{table}
  \caption{Unified Propagation Model Precision Levels}
  \centering
  \label{tab:flow}
  \begin{tabular}{ccccccccccl}
    \toprule
    Technique&Precision\\
    \midrule
    Imprecise Hardware IFT & 0 \\
    Precise Hardware IFT &  1 \\
    X-propagation &  1 \\
    Imprecise FPA & 1 \\
    Precise FPA & 2 \\
  \bottomrule
\end{tabular}
\\[10pt]
\vspace{-8mm}
\end{table}

\subsection{Synthesis for X-propagation}
\label{sec:synthesis}

\section{Unified Model Organization}
\label{sec:comparison}

Precise Hardware IFT, GLX and FPA can share the same propagation logic library from primitive gates based on their levels of precision as shown in Table \ref{tab:flow}. Here, imprecise hardware IFT is assigned a precision level of zero because it has the most false positives among different techniques. These false positives exist because neither the functionality of Boolean gates nor the effect of value on label propagation is accounted for. Precise Hardware IFT, GLX and imprecise FPA are assigned a precision level of one because they have identical propagation logic and eliminate the false positives introduced from not considering the effect of individual gates on label propagation. While level one is sufficient for Precise Hardware IFT and GLX, imprecise FPA still has false positives due to fault masking. Precise FPA is assigned a precision level of two because it eliminates false positives from not accounting for fault masking. 

\section{Experimental Results}
\label{sec:results}
\subsection{Experimental Setup}
\label{sec:setup}
In order to illustrate the difference in the levels of precision in different techniques, we perform analysis on various gate-level netlists and measure the impact of individual signals on the entire design. We implement Python scripts to parse circuit netlists into graphs where each node denotes a gate and the edges represent its input and output signals. After this, we design Python functions to perform the propagation logic for imprecise hardware IFT, Precise Hardware IFT and FPA. For each benchmark, we assign random test vectors to the input, mark different percentages of signals with a tainted or faulty label and use our propagation logic to determine whether or not these tainted or faulty signals would affect the outputs.

\subsection{Percentage Difference Analysis}
\label{sec:diff}
This section shows experimental results that quantify the difference between imprecise hardware IFT, Precise Hardware IFT and FPA propagation logic functions. First, we performed 1,000 random tests per benchmark and calculated the percentage of the tests where the tainted or faulty signals affected the output. Next, we performed t-tests and calculated bootstrap confidence intervals. T-tests were used to determine if the differences between imprecise hardware IFT and Precise Hardware IFT  were statistically significant for each benchmark and bootstrap confidence intervals were used to accurately quantify the overall percentage of difference between imprecise hardware IFT and Precise Hardware IFT for each benchmark at a 95\% confidence interval. Results are shown in Figure \ref{fig:Sig_graph_iGLIFT}.

We see that for all benchmarks except c499, c5315 and c7552 imprecise hardware IFT is always different from Precise Hardware IFT. Furthermore c499 is the only benchmark where the difference is not statistically significant. This benchmark's Precise Hardware IFT results are very similar to imprecise hardware IFT because it is constructed in such a way that tainting a decent number of signals almost always results in the output being tainted. Although the complexity of the propagation logic increases substantially, it will almost always provide considerable benefit by eliminating false positives.

T-tests and bootstrap confidence intervals were also used to specify the percentage difference between Precise Hardware IFT and FPA and the results are shown in Figure \ref{fig:Sig_graph_FPA}. We see that most benchmarks with fewer than 313 gates (c17, 74182, 74L85, 74181, and c432) show a significant difference between Precise Hardware IFT and FPA ranging from 32.3\% to 92.2 \%. Alternatively, we see that all benchmarks with more than 313 gates do not show a significant difference between Precise Hardware IFT and FPA, only being between 0.8 \% and 29.5 \% different. 
This is because the only case where GLIFT differs from FPA is fault masking. Significant differences were typical for larger circuits because faulty signals were more sparse and fault masking occurred less. However, for smaller benchmarks, differences between GLIFT and FPA are quite pronounced. Given these results, in many cases it is more practical to use precision level 1 rather than level 2 for FPA because complexity will be reduced, there will be no false negatives and there will not be a significant amount of false positives.

\section{Related Work}
\label{sec:related}
 Fault propagation, X-propagation, and IFT each have ad hoc tools which optimize the analysis of a particular technique. Fault propagation uses OneSpin Fault Injection App (FIA). X-propagation uses Synopsys VCS Xprop, Cadence IFV, and Cadence Jaspergold and Real Intent Ascent XV. IFT uses GLIFT.

FIA allows the user to define and inject fault scenarios \cite{FIA}. Fault scenarios analyze the effect on the design when particular signals are faulty and are capable of analyzing different fault types. In addition to this, FIA can associate these fault scenarios with assertions and measure the coverage of the assertions to determine which functional areas have verification gaps. 

Synopsys VCS Xprop is capable of evaluating X-values by comparing the path where the signal is set to `1' to the path where the signal is set to `0' for X-propagation \cite{XPA_Sim}. Cadence IFV, Cadence JasperGold and Real Intent Ascent XV have this capability as well along while providing their own additional features. Cadence IFV can automatically add X-propagation specific assertions, Cadence JasperGold can provide a failure trace path of a detected X-value from source to destination and Real Intent Ascent XV can interact with logic simulators to eliminate false positives at runtime. 

For IFT, GLIFT is able to track all logical information flow on Boolean gates and provide synthesizable analysis logic for creating secure architectures \cite{First_GLIFT}. In addition to this, it is possible to isolate particular areas of a design.

Although each of these tools provide unique advantages, not all of them are synthesizable and thus cannot be used in standard EDA simulation, verification, and emulation platforms. In particular, the X state is not synthesizable and is usually taken as don't care during synthesis which can be optimized to 0 or 1. Thus, it is not possible to use FPGA prototyping and emulation to accelerate X-simulation. On the other hand, our unified model provides allows designers to use FPGA prototyping and emulation platforms to accelerate these propagation problems.



\section{Conclusion}
\label{sec:conclusion}
Hardware information flow tracking, fault propagation, and X-propagation, address different problem domains. Yet, at their core they share many similarities. This paper define a unified framework to address information flow tracking, X-propagation and fault propagation using attribute labels and propagation logic at multiple levels of precision. This framework is able to take the advantage of existing hardware design tools for verification, simulation, emulation and runtime monitoring. We observe that for circuits consisting of more than 313 gates, the difference between Precise Hardware IFT and FPA is not statistically significant for ISCAS 74X-series and '85 benchmarks while the difference between imprecise hardware IFT and Precise Hardware IFT is almost always statistically significant regardless of the size of the design.

 \bibliographystyle{ACM-Reference-Format}
\bibliography{reference}

\end{document}